# Metasurface Dome for Above-the-Horizon Grating Lobes Reduction in 5G-NR Systems

Davide Ramaccia, *Senior Member, IEEE*, Mirko Barbuto, *Senior Member, IEEE*,
Alessio Monti, *Senior Member, IEEE*, Stefano Vellucci, *Member, IEEE*, Claudio Massagrande,
Alessandro Toscano, *Senior Member, IEEE*, Filiberto Bilotti, *Fellow, IEEE*

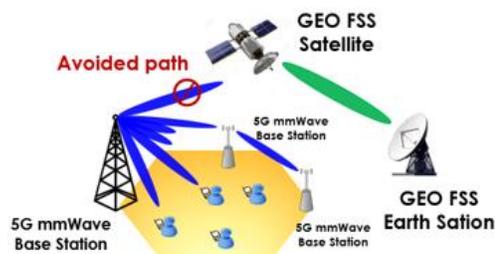

Fig. 1. Reference scenario: a 5G mm-wave base station radiates with an undesired beam above the horizon, potentially reaching the geostationary satellite operating in the adjacent band. The proposed solution aims at reducing the energy radiated on the avoided path by using a metasurface-based dome.

*Abstract*— The use of 5G New Radio (NR) spectrum around 26 GHz is currently raising the quest on its compatibility with the well-established Earth Exploration-Satellite Service (EESS), which may be blinded by the spurious radiation emitted Above-the-Horizon (AtH) by Base Station (BS) antennas. Indeed, AtH grating lobes are often present during cell scanning due to the large inter-element spacing in BS array antennas for achieving higher gains with a reduced number of RF chains. In this letter, we propose an approach based on an electrically thin metasurface-based dome for the reduction of AtH grating lobes in 5G-NR BS antennas. The proposed *scanning range shifting approach* exploits the natural lower amplitude of the grating lobes when the antenna array scans in an angular region closer to the broadside direction. The grating lobe reduction is here demonstrated considering a 1x4 phased linear antenna array operating under dual-liner ±45°-slant polarization. A simple design procedure for designing the metasurface dome is reported, together with the antenna performances, evaluated through a proper set of numerical experiments. It is shown that the grating lobe radiation towards the satellite region is significantly reduced, whereas the overall insertion loss is moderate.

*Index Terms*—5G New Radio, Angular scanning range, Metasurface, Millimeter waves, Phased Array Antenna, Radome.

## I. INTRODUCTION

THE deployment of fifth-generation (5G) cellular networks has required the expansion of the spectrum access beyond the traditional licensed spectrum bands allocated for mobile cellular networks, to make available to 5G all possible underutilized channels in the microwave and millimeter spectrum and providing coverage and gigabit connectivity [1]. However, several services already occupy such frequency bands, raising a possible interference with the 5G network service. One of the most critical frequency ranges is the one around 26 GHz, where the operative bands of 5G NR and passive satellite sensors to detect water vapor are so close that degradation of performance up to 30% is expected [2]. The reference scenario is depicted in Fig. 1. The 5G mm-wave Base station (BS) scans its Field-of-View (FoV) in both elevation and azimuth for serving the users within the cell. Since to satisfy the 5G performances highly-directive beams are needed, array antennas with an electrically large inter-element spacing are usually used. In this way, the antenna aperture is easily increased but the backend circuitry and the beam-forming network complexity is maintained unaltered [3], [4]. However, such an antenna layout leads to the raising of grating lobes pointing above-the-horizon during scanning of the elevation FoV of the antenna. The Fixed-Satellite Service (FSS) provided by a geostationary satellite may be denied due to the strong interference with such a 5G spurious radiation. The issue has been recently discussed at the International Telecommunication Union's (ITU) World Radiocommunication Conference 2019 (WRC-19) fixing the maximum emission in terms of EIRP towards satellites to 30dB(W/200 MHz) [5]. It is clear that a redesign of the antenna systems within the current 5G-NR Base stations is required for simultaneously maintaining the current radiation performances within the cell but reducing the radiation towards satellites.

The aim of this letter is to propose the *Scanning Range*

Manuscript received March 31, 2022, revised July 27, 2022.
This work has been developed in the frame of the activities of the Project MANTLES, funded by the Italian Ministry of University and Research under the PRIN 2017 Program (protocol number 2017BHFZKH). (Corresponding author: Davide Ramaccia, davide.ramaccia@uniroma3.it)
D. Ramaccia, A. Monti, S. Vellucci, A. Toscano, and F. Bilotti are with the Department of Industrial, Electronic and Mechanical Engineering, ROMA TRE University, 00146 Rome, Italy.

M. Barbuto is with the Department of Engineering, Niccolò Cusano University, 00166, Rome, Italy.
C. Massagrande is with Milan Research Center - Huawei Technologies, Milan, Italy.
Color versions of one or more figures in this communication are available at https://doi.org/XXXXXXX.
Digital Object Identifier XXXXXXXXXXX

*Shifting (SRS) approach* based on metasurfaces as a strategy for reducing the Above-the-Horizon (AtH) grating lobes radiated by standard BS array antennas layouts, and to present the design of a metasurface-based dome to be placed on top of existing antennas for implementing it.

In the state-of-the-art, several methods have been employed for reducing the grating lobes without acting on the antenna layout: virtual element filling [6], spatial hard windowing [6], optimized multilayered covers [7], or Chebyshev covers [8], just to name a few. However, they exploit the spatial filtering properties of the cover and the modulation of exciting coefficients of the antenna element, degrading the antenna matching due to reflections at the rejected angles and reducing the antenna efficiency, respectively. Rotating subarrays [9] and, more recently, rotating metasurface covers [10] have been also proposed for achieving grating lobe suppression in phased array but, here, a mechanical motion is required for optimizing the response and realizing a beam scanning within a certain FoV, making these approaches interesting, but not easily employable in pre-existing systems. Differently from the above solutions, the approach proposed here leverages the natural lower amplitude of the grating lobes when the antenna array scans in an angular region closer to broadside direction, than the one required by the coverage area. *The proposed solution aims at maximizing the backward compatibility with the legacy BS antenna systems already in use.* Here, starting from our preliminary results in [11], the design of a metadome implementing the proposed approach is presented and verified considering a dual-linear ±45°-slant polarized phased array of printed antennas with an inter-element spacing larger than the operative wavelength, emulating the typical array antenna layout used in 5G BS.

The letter is organized as follows. In Section II, the operative principle of the SRS approach is presented, together with the quantities useful for maximizing the trade-off between insertion loss in broadside direction and grating lobe reduction level. In Section III, we discuss the design of the surface property of the metadome implementing the restoration of the original scanning range based on an alignment of ideal Huygens metacells [12]. The operativity is verified through full-wave simulations. Finally, in Section IV the conclusions are drawn.

## II. SCANNING RANGE SHIFTING (SRS) APPROACH

### A. Operative principle

The operative principle of the SRS approach is described in Fig. 2. Let us consider a phased array antenna with electrically large inter-element spacing on the elevation plane. The antenna radiates a grating lobe towards the satellite region with an angle $-\theta_{GL}$, when the main lobe is pointing towards the maximum scanning angle $+\theta_{0,max}$ in the user angular region, as shown in Fig. 2a. As it is well known from antenna theory [3], reducing the maximum elevation scanning angle from $\theta_{0,max}$ to $\theta'_{0,max}$, the grating lobe moves away from the broadside direction ($-\theta'_{GL} < -\theta_{GL}$), and, simultaneously, reduces its amplitude due to the lower constructive interference of the radiated fields by the antennas (Fig. 2b before the interaction with the dome).

In the scenario in Fig. 2b, the metasurface is demanded to impart a steering to the antenna beam to restore the original scanning range below the horizon, *i.e.*, at $+\theta_{0,max}$, but benefiting of the lower grating lobes naturally reduced by the smaller scanning angular range of the antenna system. The SRS approach implemented by the metasurface-based domes modifies, thus, the operative FoV of the antenna as depicted in Fig. 2c. In elevation, the new FoV of the antenna in presence of the metadome (blue box) is *counterintuitively* shifted towards the satellite angular range of a quantity $\theta_S = |\theta'_{0,max} - \theta_{0,max}|$. According to the scanning characteristics of the antenna array and the desired grating lobe suppression level, the shifting angle $\theta_S$ in elevation must be judiciously selected and used for the design of the metasurface dome. On the contrary, the azimuthal scanning plane of 5G BS antenna is never affected by the grating lobe because of the required accuracy in user localization within the standard 120° azimuthal scanning range, *i.e.*, $[-\varphi_{0,max}, +\varphi_{0,max}] = [-60°, +60°]$. Therefore, the metasurface dome must be designed for only restoring the original scanning range in elevation but maintaining unaltered the azimuthal one, electromagnetically shifting the FoV of the antenna from the new to the original one as shown in Fig. 2c.

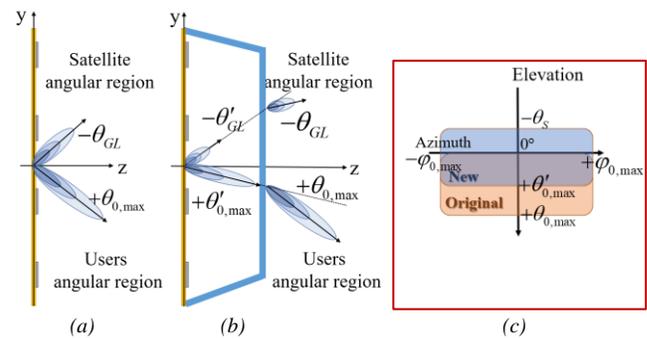

Fig. 2. Comparison between elevation radiation patterns of the antenna (a) without and (b) with the metasurface dome. (c) Operative FoVs in elevation and azimuth when the SRS approach is applied (new FoV, blue scanning area) and not applied (original FoV, orange scanning area).

### B. SRS approach applied to a phased array antenna

In this section, we describe the quantities to be used for identifying the best trade-off between broadside insertion loss and grating lobe reduction level when the SRS approach is applied to a generic phased array antenna, emulating the typical array antenna layout used in 5G BS.

Let us consider the phased array shown in Fig. 3a. The antenna consists of four square patch antennas operating at the frequency $f_0$, each of which is rotated at 45° for implementing the typical dual-linear ±45°-slant polarization of BS antennas. The inter-element spacing along y-direction (elevation plane) is set to $d = 1.2\lambda_0$, *i.e.*, larger than the operative free-space wavelength $\lambda_0 = c/f_0$, whereas the width $w$ along x-direction is half of the operative wavelength. As typically done in sparse array antennas, the antenna elements are engineered to exhibit a null in the direction of the array factor grating lobe when pointing in the broadside direction, *i.e*, close to the directions ±60°. However, this strategy is no more effective when the array points out of the broadside direction. In Fig. 3b, the co- and cross-polar gains as a function of the elevation angle are reported for three different pointing directions, *i.e.*, $\theta_0 = 0°, 10°, 20°$. The numerical results have been carried out using CST MS [13] and using open boundaries in $y$-/$z$-directions and periodic boundary conditions along $x$-direction. Scanning the elevation plane in the user angular region, the grating lobes (GL) are clearly present for negative angles, *i.e.*, satellite angular region, even for small scanning angles for both the co- and cross-polar gains. Neglecting the cross-polar components (dashed lines in Fig, 3b), which are always 20 dB lower than the co-polar ones, we highlight that the co-polar GL level is higher than -10 dB (-9.9dB in Fig. 3b) already for $\theta_0 = 10°$, and rapidly increase reaching -2.45dB for $\theta_0 = 20°$, leading to a strong radiation towards FSS satellites.

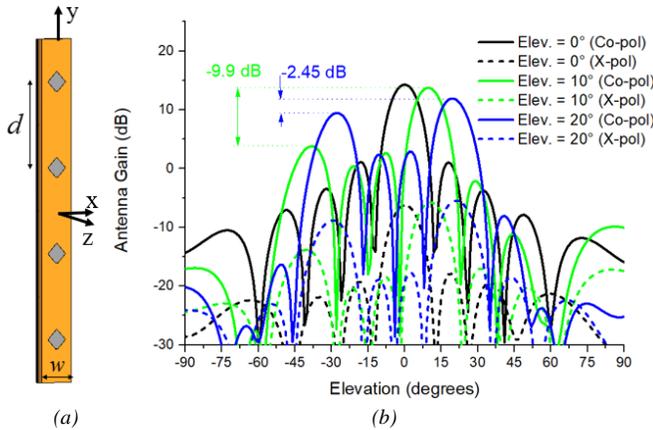

Fig. 3. (a) Array antenna composed by 4 dual-liner 45°-slant polarized patch antenna operating at the frequency $f_0$; (b) Elevation co-polar gain for different pointing directions within the Field of View 0°-20°.

To identify the best angular shifting value $\theta_s$ to be used for designing the metadome, we set the maximum elevation angle $\theta_{0,max} = 20°$ and consider the following three key *parameters*:

1. Virtual Insertion Loss (IL) in broadside direction.
2. Gain enhancement for pointing at max elevation angle $\theta_{0,max}$
3. GL reduction for pointing at max elevation angle $\theta_{0,max}$

*Parameter no. 1* describes the best achievable insertion loss when the antenna-metadome system is radiating in broadside direction, assuming that the metadome is full-transmitting with 100% efficiency. Indeed, these "losses" are not due to the metadome, but to the scan loss of the array, given by the single antenna element radiation pattern, since it is actually radiating towards the negative pointing direction $-\theta_s$. *Parameter no. 2* represents the gain enhancement due to lower scan loss of the array antenna when the antenna+metadome system is pointing towards $\theta_{0,max}$. Finally, *parameter no. 3* represents the minimum achievable GL reduction, which corresponds to the difference between the GL of the array when points towards

$\theta_{0,max}$ and the GL observed when points towards $\theta_{0,max} - \theta_s$. The parameters are defined assuming small shifting angle $\theta_s$.

In Table I, we report the numerically computed parameters for the array without the metadome as shown in Fig. 3a, considering five different shifting angles $\theta_s$ ranging from 5° to 15°, with a step of 2.5°. As expected, the virtual IL in broadside direction increases due to the reduced aperture observed in broadside direction by the array for larger shifting angles. To balance this virtual loss, there is the gain enhancement at the max elevation scanning angle $\theta_{0,max} = 20°$. Finally, the GL reduction in satellite angular region (see Fig. 2) increases for wider shifting angles $\theta_s$, as expected.

To verify the SRS approach proposed here, we select $\theta_s = 12.5°$ from Table I, which ensures the best trade-off performances in terms of GL reduction and Virtual IL. This value is used in the next section for designing the metadome.

## III. METASURFACE-BASED DOME

The SRS approach can be implemented using a linear gradient refracting metasurface [12], [14]–[22], able to impart a wavevector shift to the beams radiated by the antenna, Fig. 4(a). Several works can be found on the design of refractive metasurfaces. Among them, it is worth mentioning the analysis and design methods based on a cascade of equivalent circuit matrices [21],[17], or on the polarizability of the particles composing the metasurface [18].

Here, for sake of completeness, we briefly summarize the design procedure for deriving the required spatial profile of the metasurface dome for implementing the proposed SRS approach for $\theta_s = 12.5°$. In Fig. 4a, the incident wave with $\mathbf{k}_{0,i} = k_{y,i}\mathbf{y} + k_{z,i}\mathbf{z}$ is illuminating the metasurface with an angle $\theta'_{0,max}$. It must be transformed into a new wave with wavevector $\mathbf{k}_{0,t} = k_{y,t}\mathbf{y} + k_{z,t}\mathbf{z}$ propagating towards $\theta_{0,max}$. Being the material on the two sides of the metasurface vacuum ($\varepsilon_0, \mu_0$), both the incident and refracted wave must have the same wavenumber $k_0 = 2\pi/\lambda_0$. Therefore, the anomalous refraction of the transmitted wave towards the user region is achievable only if the metasurface imparts an additional momentum to the incident wave along the $y$-direction. The incident/transmitted wavevector components are, thus, written as follows:

$$k_{0,i} = k_{0,t}; \quad k_{y,t} = k_{y,i} + k_{MTS}; \quad k_{z,t}^2 = k_{0,t}^2 - k_{y,t}^2 \quad (1)$$

where the subscripts $i$, $t$ identify incident and transmitted waves, respectively, $k_{MTS} = 2\pi/p_{MTS}$ is the additional tangential momentum imparted by the metasurface for achieving the desired steering, and $p_{MTS}$ is the period along $y$-direction of the phase of the transmission coefficient. Imposing $k_{MTS} = k_{y,t} - k_{y,i}$ and using (1), we obtain:

$$k_{MTS} = k_0 \left| \sin\theta'_{0,max} - \sin\theta_{0,max} \right|. \quad (2)$$

from which we can derive the metasurface periodicity of the transmission phase profile for $\theta'_{0,max} = 7.5°$ and $\theta_{0,max} = 20°$ as follows:

| Shifting angle $\theta_s$ | Virtual IL Broadside | Max ELE Gain Enhanc. | GL reduction |
|---|---|---|---|
| 5.0° | 0.22 dB | 0.9 dB | 1.87 dB |
| 7.5° | 0.45 dB | 1.50 dB | 4.28 dB |
| 10.0° | 0.50 dB | 1.8 dB | 5.67 dB |
| **12.5°** | **0.86 dB** | **2.0 dB** | **7.29 dB** |
| 15.0° | 1.35 dB | 2.2 dB | 10.56 dB |

.

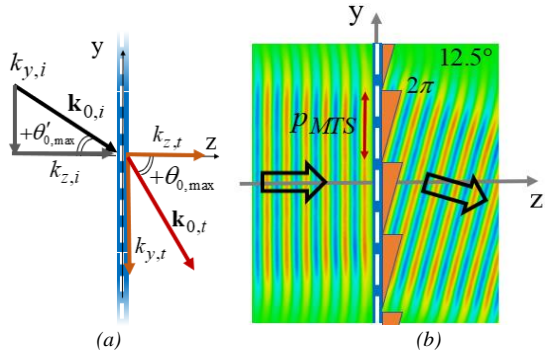

Fig. 4. (a) Wavevector components of the waves before and after the steering metasurface dome; (b) Numerically computed e-field map of a wave passing through the refracting metasurface with $p_{MTS} = 4.73\lambda_0$.

TABLE II
METADOME UNIT-CELL PERFORMANCES AND VALUES

| Cell no. | Transmission (Mag. Phase) | $Y_1$ @ $f_0$ ($\Omega$/sq.) | $Y_2$ @ $f_0$ ($\Omega$/sq.) | $Y_3$ @ $f_0$ ($\Omega$/sq.) |
|---|---|---|---|---|
| 1 | (0.98, 15°) | -j 270 | -j 48 | -j 270 |
| 2 | (0.99, 45°) | -j 141 | -j 130 | -j 141 |
| 3 | (0.97, 75°) | -j 241 | -j 100 | -j 241 |
| 4 | (0.99, 105°) | -j 366 | -j 100 | -j 366 |
| 5 | (0.96, 135°) | -j 557 | -j 129 | -j 557 |
| 6 | (0.98, 165°) | -j 957 | -j 270 | -j 957 |
| 7 | (0.99, 195°) | -j 2888 | -j 1085 | -j 2888 |
| 8 | (0.96, 225°) | +j 2479 | +j 203 | +j 2479 |
| 9 | (0.95, 255°) | +j 746 | +j 138 | +j 746 |
| 10 | (0.99, 285°) | +j 362 | +j 138 | +j 362 |
| 11 | (0.99, 315°) | +j 175 | +j 203 | +j 175 |
| 12 | (0.98, 345°) | +j 51 | +j 1084 | +j 51 |

$$p_{MTS} = \lambda_0 / \left| \sin\theta'_{0,\max} - \sin\theta_{0,\max} \right| \approx 4.73\lambda_0. \quad (3)$$

A refracting metasurface, consisting of a cascade of three reactive sheets, is used for both verifying the steering capability of the system and the SRS approach based on it (see Sec. IV for application to the antenna array). The periodicity of the metasurface $p_{MTS}$ has been sampled by 12 unit-cells, whose dimensions are $w = 0.5\lambda_0$, $l = p_{MTS}/12 \approx 0.394\lambda_0$, and reported in Fig. 5a. To cover the entire $2\pi$ -coverage within $p_{MTS}$, each cell must exhibit a transmission phase higher of 30° than the adjacent cell. Following the design strategy reported in [21], each cell consists of three reactive sheets modelled through their surface admittances $Y_1, Y_2, Y_3$, separated by two low-loss dielectric spacers of Rogers RO3003 ( $\varepsilon_r = 3.3$ , $\tan\delta = 0.001$ ) of thickness $t = 0.132\lambda_0$. In Table II, we report the metadome unit-cell performances and the values of the surface admittances. The steering capability has been verified in CST Microwave studio by illuminating about 5 metasurface periods with a plane wave. The e-field map is reported in Fig. 4(b), demonstrating the almost perfect anomalous refraction introduced by the metasurface.

## IV. 5G NR ANTENNA WITH SRS META-DOME: RESULTS AND DISCUSSION

In this section, we apply the proposed approach on a scaled version of the 5G NR phased array antenna, consisting of dual-linear ±45°-slant polarized phased array of printed antennas with an inter-element spacing larger than the operative wavelength, emulating the array antenna layout used in 5G BS.

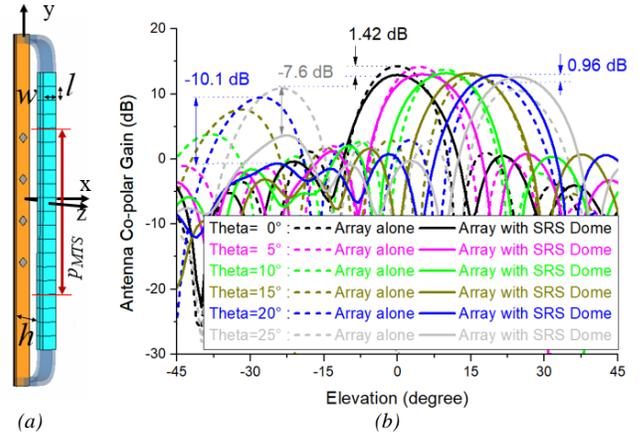

Fig. 5. (a) Array antenna with SRS metadome, composed by a liner arrangement of the unit-cells in Table II; (b) Comparison of radiation performances between the antenna with (solid lines) and without (dashed lines) SRS metadome for the relevant pointing directions, within the angular range $\theta_0=[0°;20°]$ and also for out-of-range scanning direction $\theta_0=25°$.

The antenna array covered by the SRS metadome is shown in Fig. 5a. The metadome consists of 20 unit-cells (12 covering the required $2\pi$ -coverage within the metasurface period $p_{MTS}$ and further 8 cells that extends its active area for ensuring the operation when the array is scanning within its new FoV, Fig. 2c) and two dielectric supports made of Rogers RO3003 connected to the array. The distance between the array and the metadome is $h = \lambda_0$. In Fig. 5b, we report the numerically computed co-polar gains exhibited by the antenna array with and without the SRS metadome for the most relevant pointing directions, *i.e.*, broadside direction ($\theta_0=0°$) and maximum elevation scanning ($\theta_0=20°$). The obtained GL reduction level in satellite angular region (10.1dB), insertion loss for broadside pointing direction (1.42dB), and gain enhancement at $\theta_0=20°$ (0.96 dB) are reported, and in line with the values discussed in Sec. IIb. The GL reduction performances are also good out of the angular range of interest (-7.6dB at $\theta_0=25°$), showing that the metasurface radome can operate even beyond the designed angular range. The results demonstrate the validity of the approach that leverages on the natural lower grating lobes when the antenna array scans in an angular region closer to the broadside direction, than the one required by the coverage area.

## V. CONCLUSION

In this letter, we have proposed the *SRS approach* based on metadome for reducing the AtH GLs radiated by BS array antennas. We have discussed the operative principle, and derived the application strategy considering a scaled version of the 5G NR phased array with an inter-element spacing larger than the operative wavelength, for emulating the typical array layout used in 5G BS. The design of a SRS metadome based on a refracting Huygens metasurface able to impart a constant wavevector shift for changing the operating FoV of the antenna has been reported. The full-wave numerical simulations demonstrate the feasibility of the approach and the validity of the performance analysis through the three key parameters used for selecting the best scanning range shifting angle.

REFERENCES

[1] J. G. Andrews *et al.*, "What will 5G be?," *IEEE J. Sel. Areas Commun.*, vol. 32, no. 6, pp. 1065–1082, 2014.

[2] J. Behrens, "NOAA Warns 5G Spectrum Interference Presents Major Threat to Weather Forecasts," *American Institute of Physics*, 2019.

[3] C. A. Balanis, *Antenna theory : analysis and design*. Wiley, 2005.

[4] J. S. Herd and M. David Conway, "The Evolution to Modern Phased Array Architectures," *Proc. IEEE*, vol. 104, no. 3, pp. 519–529, 2016.

[5] ITU, "Terrestrial component of International Mobile Telecommunications in the frequency band 24.25-27.5 GHz (RES 242)," 2019.

[6] P. Chakravorty and D. Mandal, "Grating Lobe Suppression With Discrete Dipole Element Antenna Arrays," *IEEE Antennas Wirel. Propag. Lett.*, vol. 15, pp. 1234–1237, 2016.

[7] Y. J. Lee, S. H. Jeong, W. S. Park, J. S. Yun, and S. I. Jeon, "Multilayer spatial angular filter with airgap tuners to suppress grating lobes of $4 \times 1$ array antenna," *Electron. Lett.*, vol. 39, no. 1, pp. 15–17, Jan. 2003.

[8] R. Mailloux, "Synthesis of Spatial Filters with Chebyshev Characteristics," *IEEE Trans. Antennas Propag.*, vol. 24, no. 2, pp. 174–181, 1976.

[9] V. D. AGRAWAL, "Grating-Lobe Suppression in Phased Arrays by Subarray Rotation," *Proc. IEEE*, vol. 66, no. 3, pp. 347–349, 1978.

[10] K. Singh, M. U. Afzal, M. Kovaleva, and K. P. Esselle, "Controlling the Most Significant Grating Lobes in Two-Dimensional Beam-Steering Systems with Phase-Gradient Metasurfaces," *IEEE Trans. Antennas Propag.*, vol. 68, no. 3, pp. 1389–1401, 2020.

[11] A. Monti *et al.*, "Gradient Metasurface Dome for Phased arrays able Reducing the Grating Lobes within Single-side Scanning region," in *2021 IEEE International Symposium on Antennas and Propagation and USNC-URSI Radio Science Meeting (APS/URSI)*, 2022, pp. 729–730.

[12] C. Pfeiffer and A. Grbic, "Metamaterial Huygens' Surfaces: Tailoring Wave Fronts with Reflectionless Sheets," *Phys. Rev. Lett.*, vol. 110, no. 19, p. 197401, May 2013.

[13] "CST Studio Suite 3D EM simulation and analysis software." [Online]. Available: https://www.3ds.com/products-services/simulia/products/cst-studio-suite/?utm_source=cst.com&utm_medium=301&utm_campaign=cst. [Accessed: 05-Apr-2020].

[14] F. Monticone, N. M. Estakhri, and A. Alù, "Full Control of Nanoscale Optical Transmission with a Composite Metascreen."

[15] C. Pfeiffer and A. Grbic, "Millimeter-wave transmitarrays for wavefront and polarization control," *IEEE Trans. Microw. Theory Tech.*, vol. 61, no. 12, pp. 4407–4417, Dec. 2013.

[16] A. Epstein and G. V. Eleftheriades, "Passive Lossless Huygens Metasurfaces for Conversion of Arbitrary Source Field to Directive Radiation," *IEEE Trans.*

*Antennas Propag.*, vol. 62, no. 11, pp. 5680–5695, Nov. 2014.

[17] J. P. S. Wong, A. Epstein, and G. V. Eleftheriades, "Reflectionless Wide-Angle Refracting Metasurfaces," *IEEE Antennas Wirel. Propag. Lett.*, vol. 15, pp. 1293–1296, 2016.

[18] V. S. Asadchy, M. Albooyeh, S. N. Tcvetkova, A. Díaz-Rubio, Y. Ra'di, and S. A. Tretyakov, "Perfect control of reflection and refraction using spatially dispersive metasurfaces," *Phys. Rev. B*, vol. 94, no. 7, p. 075142, Aug. 2016.

[19] K. Chen *et al.*, "A Reconfigurable Active Huygens' Metalens," *Adv. Mater.*, vol. 29, no. 17, p. 1606422, May 2017.

[20] A. Díaz-Rubio, V. S. Asadchy, A. Elsakka, and S. A. Tretyakov, "From the generalized reflection law to the realization of perfect anomalous reflectors," *Sci. Adv.*, vol. 3, no. 8, 2017.

[21] M. Chen, E. Abdo-Sánchez, A. Epstein, and G. V. Eleftheriades, "Theory, design, and experimental verification of a reflectionless bianisotropic Huygens' metasurface for wide-angle refraction," *Phys. Rev. B*, vol. 97, no. 12, p. 125433, Mar. 2018.

[22] V. G. Ataloglou and G. V. Eleftheriades, "Arbitrary Wave Transformations with Huygens' Metasurfaces through Surface-Wave Optimization," pp. 1–5, 2021.